\documentclass[letter]{aa}  
\usepackage{comment}
\usepackage{amsmath}    
\usepackage{amssymb}    
\usepackage{upgreek}
\usepackage{tikz}
\usetikzlibrary{matrix}
\usepackage{graphicx}
\usepackage{bm}
\usepackage{multirow}
\usepackage[varg]{txfonts}
\usepackage{natbib}
\bibpunct{(}{)}{;}{a}{}{,} 
\usepackage[colorlinks=true,linkcolor=blue,citecolor=blue]{hyperref}%
\usepackage{orcidlink}
\newcommand{\ha}{H$\alpha$}

\newcommand{\hb}{H$\beta$}

\newcommand{\chandra}{{\it Chandra}}

\def\aap{A\&A}

\def\apjl{ApJL}
\def\apjs{ApJS}

\newcommand{\oiii}{[O{\,\sc iii}]}

\newcommand{\feii}{Fe\,{\sc ii}}

\usepackage{graphicx}
\usepackage{txfonts}
\begin{document} 

   \title{The changing-look AGN SDSS J101152.98+544206.4 is returning to a type I state}
  \titlerunning{SDSS J101152.98+544206.4}
  

   \author{Bing Lyu \orcidlink{0000-0001-8879-368X}
          \inst{1} \fnmsep \thanks{}
          \and
          Xue-Bing Wu \orcidlink{0000-0002-7350-6913}
          \inst{1,2} \fnmsep \thanks{}
           \and        
          Yuxuan Pang \orcidlink{0009-0005-3823-9302} \inst{1,2} 
          \and  
          Huimei Wang \orcidlink{0000-0001-8803-0738} \inst{1,2} 
          \and  
          Rui Zhu \orcidlink{0000-0002-0792-2353} \inst{1,2} 
          \and  
          Yuming Fu\orcidlink{0000-0002-0759-0504} \inst{1,2,3,4}  
          \and  
          Qingwen Wu \orcidlink{0000-0003-4773-4987} \inst{5}
          \and
          Zhen Yan \orcidlink{0000-0002-5385-9586}\inst{6}
          \and          
          Wenfei Yu \orcidlink{0000-0002-3844-9677}\inst{6}
           \and
          Hao Liu \orcidlink{0000-0001-5525-0400} \inst{7}   
          \and
          Shi-Ju Kang \orcidlink{0000-0002-9071-5469} \inst{8}   
          \and
          Junjie Jin \orcidlink{0000-0002-8402-3722} \inst{9}  
          \and
          Jinyi Yang \orcidlink{0000-0001-5287-4242} \inst{10}  
          \and
          Feige Wang \orcidlink{0000-0002-7633-431X} \inst{10}           
          }
   \institute{Kavli Institute for Astronomy and Astrophysics, Peking University, Beijing 100871, People's Republic of China\\
              \email{lyubing@pku.edu.cn}          
           \and
        Department of Astronomy, School of Physics, Peking University, Beijing, 100871, People's Republic of China \\
             \email{wuxb@pku.edu.cn}
             \and 
             Leiden Observatory, Leiden University, P.O. Box 9513, NL-2300 RA Leiden, The Netherlands 
             \and 
             Kapteyn Astronomical Institute, University of Groningen, P.O. Box 800, NL-9700 AV Groningen, The Netherlands           
              \and
             Department of Astronomy, School of Physics, Huazhong University of Science and Technology,
1037 Luoyu Road, Wuhan, 430074, People's Republic of China             
          \and
            Shanghai Astronomical Observatory, Chinese Academy of Sciences, 80 Nandan Road,
Shanghai, 200030, People's Republic of China            
              \and
              University of Science and Technology of China,
No.96, JinZhai Road Baohe District, Hefei, Anhui, 230026, People's Republic of China 
             \and
             School of Physics and Electrical Engineering,  Liupanshui Normal University,  Liupanshui, Guizhou, 553004, People's Republic of China
             \and
             Key Laboratory of Optical Astronomy, National Astronomical Observatories, Chinese Academy of Sciences, Beijing 100012, People's Republic of China
             \and 
             Steward Observatory, University of Arizona, 933 N Cherry Avenue, Tucson, AZ 85721, USA
             }

   \date{Received  , 2024; accepted  , 2024}

 
  \abstract
   {} 
   {We discovered that the changing-look active galactic nucleus (CLAGN) SDSS J101152.98+544206.4 (J1011+5442 for short) gradually returns to the type I state after a short period between 2014 and 2019 in the faint type 1.9 state.
   }
   {Motivated by the rebrightening in the optical and mid-infrared light curves from ZTF and WISE, we obtained new spectroscopic observations with the Xinglong 2.16 m, the Lijiang 2.4 m, and the MMT 6.5 m optical telescopes in 2024. }
   {After changing its optical AGN type from 1 to 1.9 between 2003 and 2015 according to the repeat spectroscopy from the Time Domain Spectroscopic Survey, J1011+5442 returned to its type I state in 2024. We detect the significant and very broad \hb\, lines (full width at half maximum of $ \gtrsim 5000\, {\rm km/s } $) based on the new spectra, which suggests that J1011+5442 was in the intermediate state between the dim state in 2015 and the bright state in 2003. The long-term optical and mid-infrared light curves also show a brightening trend between 2019 and 2024 as the broad \hb\, line appeared. The time lag of about 100 days between the mid-infrared and optical variability is consistent with the prediction of dust reverberation mapping.  
   }
   {The behavior of the photometric and spectroscopic observations of J1011+5442 is consistent with the argument that the repeating changing-look phenomenon is regulated by the variation in the accretion rate. }



   \maketitle
%
\nolinenumbers

\section{Introduction} \label{sec:intro}
A so-called changing-look active galactic nucleus (CLAGN) is a special type of AGNs that exhibits a significant change in their optical spectral appearance. CLAGNs experience transitions from a type I AGN with both broad (e.g., full width at half maximum, $\rm FWHM  \gtrsim 1000\,km/s$) and narrow emission lines to a type II AGN with only narrow lines ($\rm FWHM \lesssim 500\,km/s$), or vice versa \citep[e.g.,][]{2023NatAs...7.1282R}. The classification of intermediate types 1.5, 1.8, and 1.9 are also defined based on the relative strength of the broad \hb\, lines to the narrow \oiii\, lines \citep[][]{1976MNRAS.176P..61O,1981ApJ...249..462O,1992MNRAS.257..677W}. Increasingly more optical CLAGNs are found based on various methods \citep[e.g.,][]{2019ApJ...874....8M,2020ApJ...889...46S,2021A&A...650A..33P,2021MNRAS.503.2583S,2022ApJ...926..184J,2023ApJ...953...61Y,2024ApJS..270...26G}. The main plausible mechanism for CLAGNs is related to (1) the variation in the intrinsic accretion rate \citep[e.g.][]{2017ApJ...846L...7S,2018MNRAS.480.3898N,2021MNRAS.508..144G,2022ApJ...930...46L,2023ApJ...953...61Y}, (2) obscuration effects \citep[e.g.,][]{2009MNRAS.393L...1R,2016ApJ...820....5R}, and (3) tidal disruption events \citep[TDEs, e.g.,][]{2019ApJ...883...94T}. The obscuration scenario is disfavored for most sources with low absorption, low polarization degrees, and significant multiwavelength variability. TDE can only be applied to limited cases, and it is hard to explain the repeating changing-look phenomena. CLAGNs that show repeating appearance/disappearance of broad emission lines are much rarer (e.g., \citealp[3C 390.3, ][]{2019sf2a.conf..509M}; \citealp[B3 0749+460A, ][]{2022RAA....22a5011W}; \citealp[Fairall 9, ][]{2019sf2a.conf..509M}; \citealp[Mrk 1018, ][]{2016A&A...593L...8M}; \citealp[Mrk 590, ][]{2014ApJ...796..134D}; \citealp[NGC 1365, ][]{2014ApJ...795...87B,2023MNRAS.518.2938T}; \citealp[NGC 1566, ][]{2020A&A...641A.167S}; \citealp[NGC 4151, ][]{2007MNRAS.377..607P,2019sf2a.conf..509M}; \citealp[NGC 5548, ][]{2020A&A...641A.167S};  \citealp[NGC 7582, ][]{2009ApJ...695..781B,2019sf2a.conf..509M}; \citealp[NGC 7603, ][]{2019sf2a.conf..509M}; \citealp[SDSS J151652.48+295413.4, ][]{2023ApJ...956..137W}; \citealp[UGC 3223, ][etc]{2020ApJ...901....1W}). It is still uncertain whether every CLAGN experiences repeating changing-look events, and the physical mechanisms that cause these events are unclear as well. In addition, the short changing-look timescale (e.g., several years or shorter) is still unclear \citep[e.g.,][]{2018MNRAS.480.3898N,2018ApJ...864...27S,2018MNRAS.480.4468R,2019MNRAS.483L..17D,2021ApJ...916...61F,2022AN....34310065S}. The transition between a standard Shakura-Sunyaev thin disk \citep[SSD; ][]{1973A&A....24..337S} and advection-dominated accretion flow \citep[ADAF; e.g.,][]{2008ARA&A..46..475H,2014ARA&A..52..529Y} caused by the radiation pressure instability in a narrow unstable zone might explain the multiple changing-look phenomena \citep{2020A&A...641A.167S}.  The large-scale magnetic fields can reduce the timescale for repeating CLAGNs that are driven by the wind, which can effectively remove the angular momentum and energy of the disk \citep{2021ApJ...910...97P}. 
In addition, the repeating changing-look phenomena might also be explained by the recoiled supermassive black hole (rSMBH) scenario \citep[e.g.,][]{2018ApJ...861...51K} or throuhg close supermassive black hole binaries \citep[CB-SMBHs; e.g.,][]{2020A&A...643L...9W}. The discovery of more repeating CLAGNs with high-cadence photometric and multi-epoch spectroscopic observations may help to solve this issue.

The source SDSS J101152.98+544206.4 (J1011+5442, hereafter) at $z = $ 0.246 was initially identified as a type I AGN with broad Balmer emission lines in 2003. The broad $\rm H\beta$ emission lines subsequently disappeared before 2015 \citep[][]{2016MNRAS.455.1691R}. Additionally, it exhibits a gradual decrease in its continuum luminosity. The polarization degree of J1011+5442 is low \citep[0.15 \%; ][]{2017A&A...604L...3H,2019A&A...625A..54H}. Inspired by its rebrightening since 2018 in optical and mid-infrared light curves from the \textit{Zwicky} Transient Facility \citep[ZTF; e.g.,][]{2019PASP..131a8002B} and the {Wide-field Infrared Survey Explorer} \citep[WISE;][]{2010AJ....140.1868W}, we performed follow-up spectroscopic observations with the Xinglong 2.16 m (XLT for short), the Lijiang 2.4 m (LJT for short), and the Multiple mirror telescope (MMT for short) 6.5 m optical telescopes. J1011+5442 almost returned to the type I state as the broad $\rm H\beta$ emission line reappeared. The new data confirm that J1011+5442 is a new repeating CLAGN with gradually brightened broad emission lines and continuum after 5 years in the faint state. Throughout this work, we adopt a flat $\Lambda$CDM cosmological model with $H_0$=70 km s$^{-1}$ Mpc $^{-1}$, $\Omega_{m}$=0.27, and $\Omega_{\Lambda}=0.73$, which corresponds to a luminosity distance of $D_L = 1.244 $ Gpc.  

\begin{figure*}
\centering
\includegraphics[width=0.7\textwidth]{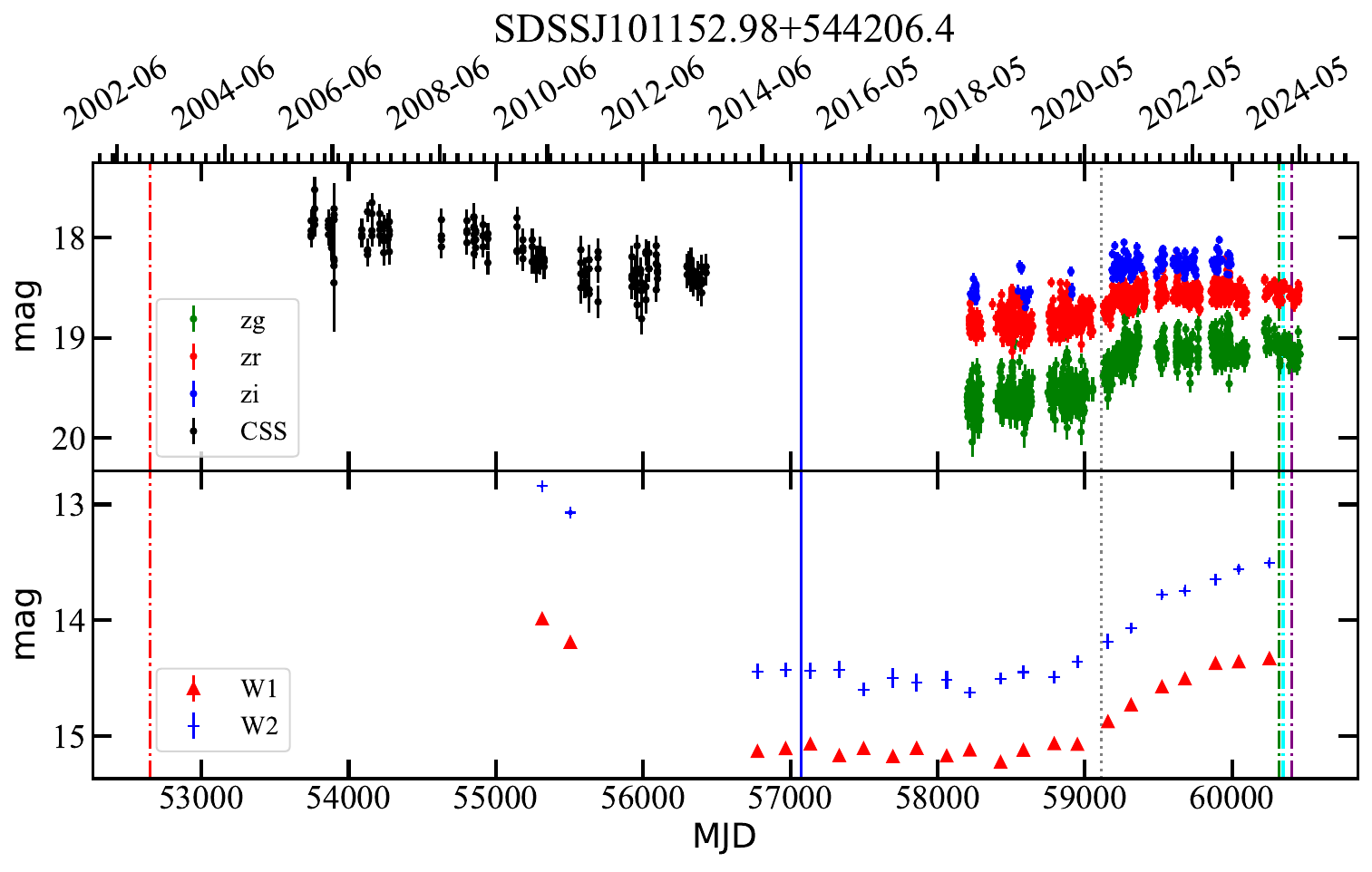}
\caption{Optical and mid-infrared light curves of J1011+5442. The dot-dashed red, green, cyan, and purple lines represent the spectroscopic observational time by SDSS, the Xinglong 2.16 m, the Lijiang 2.4 m telescope, and the MMT 6.5 m telescope. The straight blue line represents the spectroscopic observational time in the faint state by SDSS. The dotted black line represents the observational time by \chandra\,. The mid-infrared light curve is rebinned within half a year.}
\label{pic:lc}
\end{figure*}

\begin{figure*}
\centering
\includegraphics[width=0.7\textwidth]{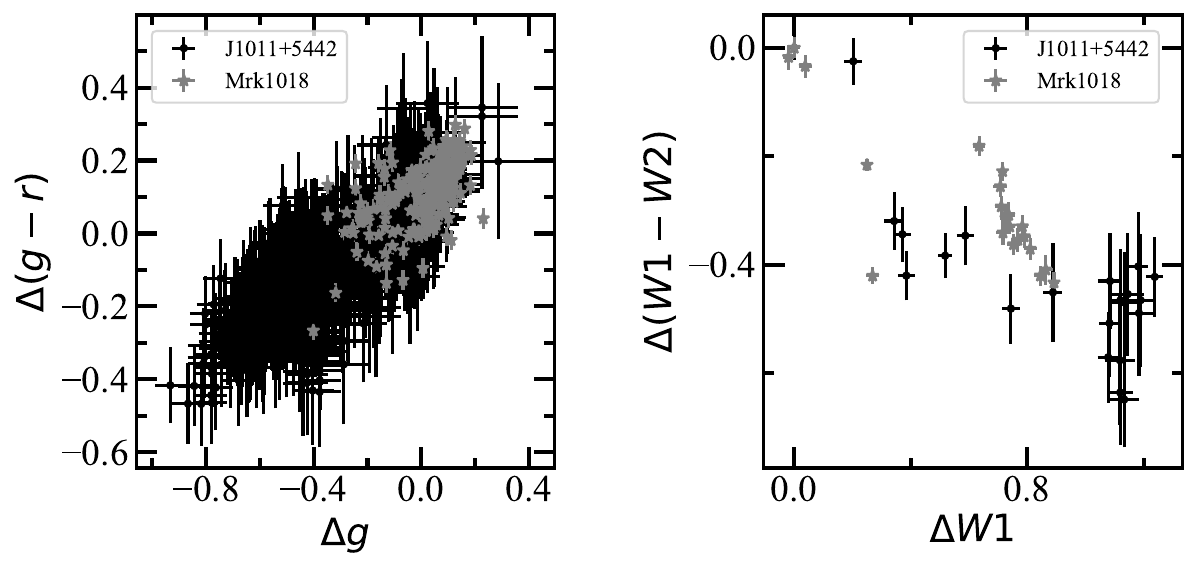}
\caption{ Color variability vs. magnitude variability of J1011+5442 in the optical (left panel) and mid-infrared (right panel) band. We include the data of Mrk 1018 for comparison. A bluer-when-brighter feature in the optical and a redder-when-brighter feature in the mid-infrared are confirmed and consistent with other CLAGNs \citep[e.g.,][]{2018ApJ...862..109Y}.
}
\label{pic:var}
\end{figure*}

\begin{figure}
\centering
\includegraphics[width=0.48\textwidth]{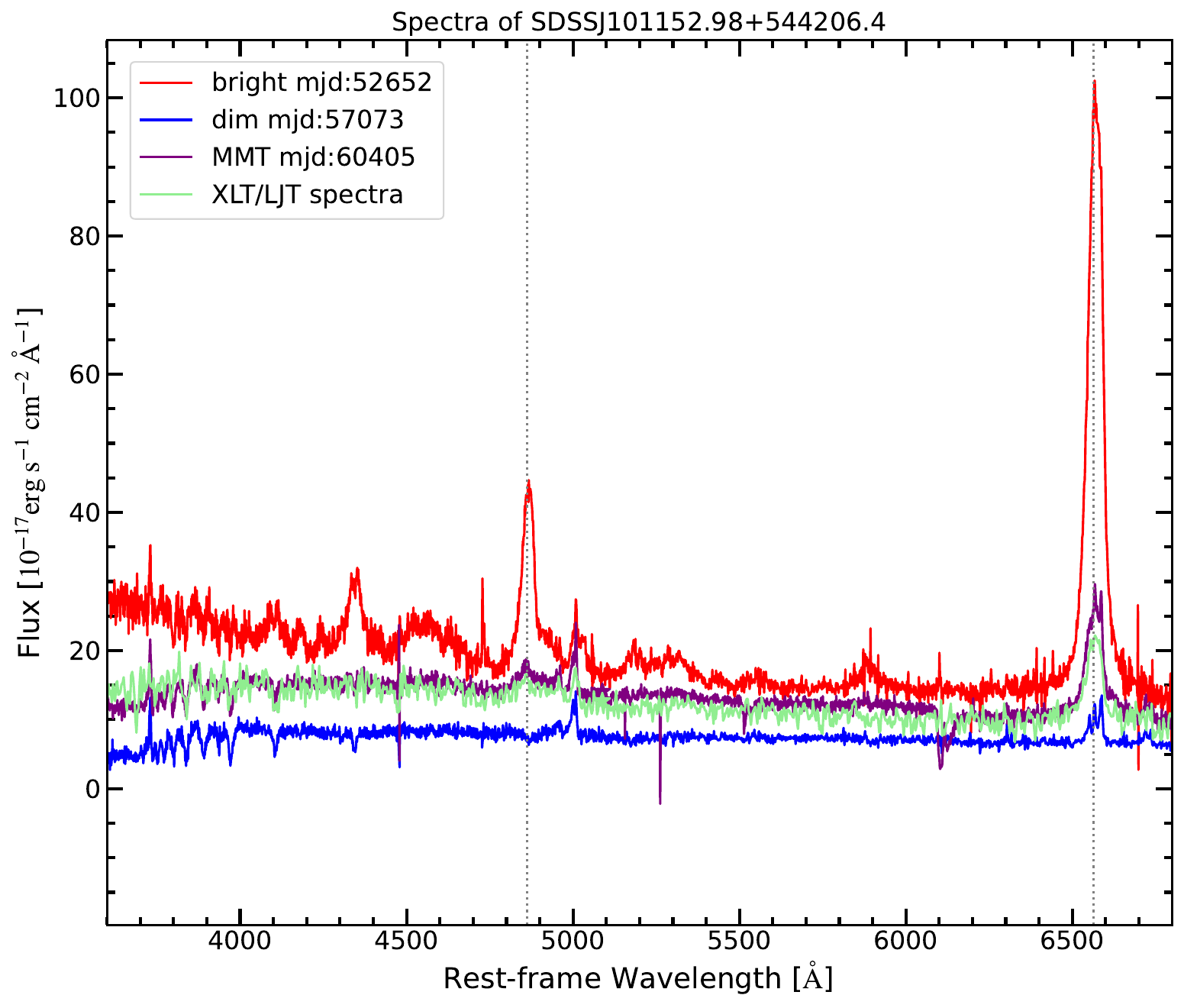}
\caption{Optical spectra of J1011+5442 in the rest frame. The bright and dim state spectra in red and blue are from the SDSS.  For the XLT/LJT and MMT spectra in light green and purple, the absolute flux is scaled with the ZTF magnitude. } 
\label{pic:spec_ztf}
\end{figure}

\begin{figure*}
\centering
\includegraphics[width=0.7\textwidth]{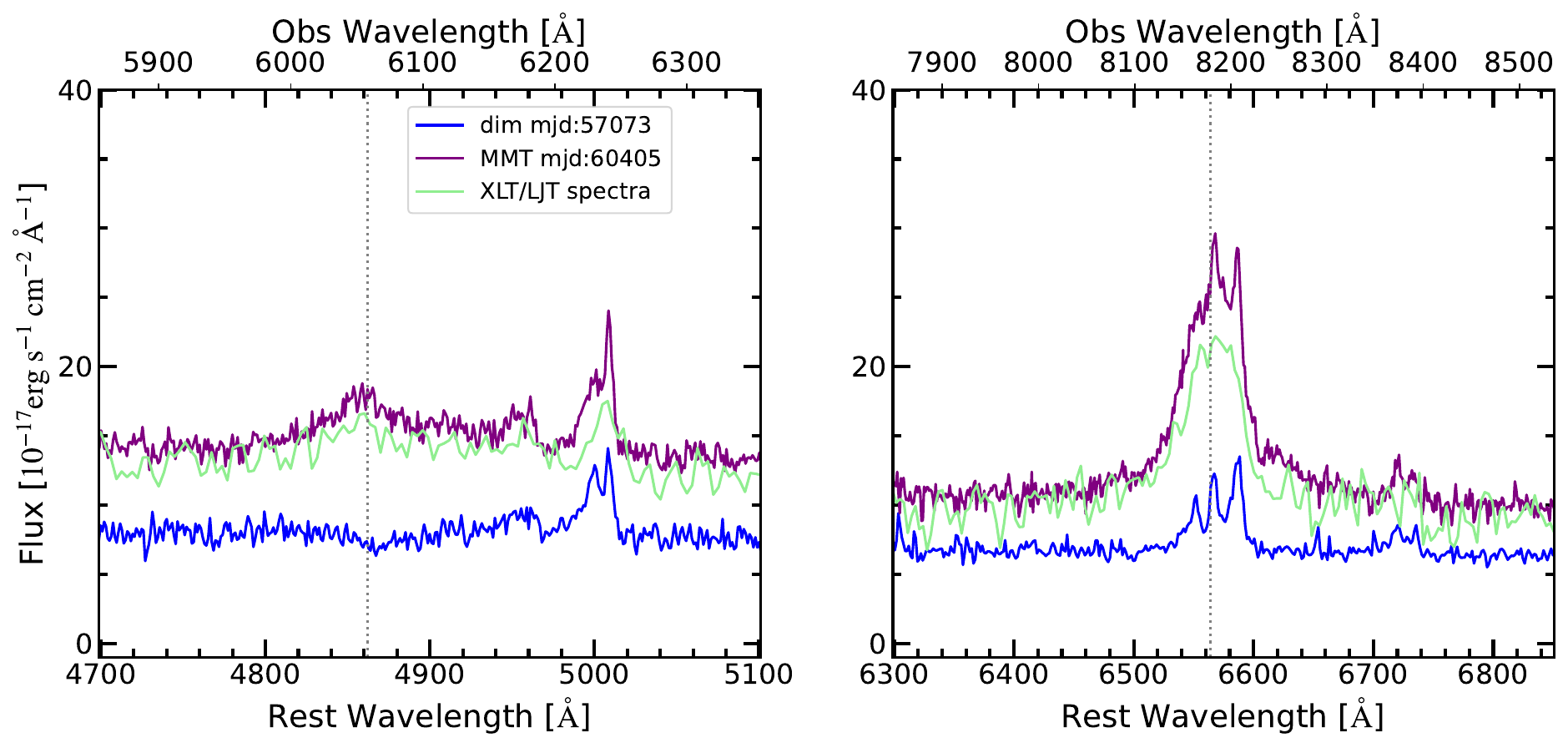}
\caption{ZTF magnitude-scaled spectra from new observations focused on the \hb\, and \ha\, emission lines. The dim-state spectrum is also shown for comparison. Broad \hb\, lines are apparent in the new spectra. }
\label{pic:hbha}
\end{figure*}

\section{Observations and data analysis} \label{sec:data}
\subsection{Photometry} \label{sec:photo}
Between 2006 January and 2013 May, J1011+5442 was observed by the Catalina Sky Survey \citep[CSS; ][]{2009ApJ...696..870D}. The ZTF has been monitoring J1011+5442 in the g, r, and i bands since 2018. We found that it gradually brightened since 2018 and peaked at around 2021, then experienced strong short-term variability in the optical light curves. We also checked the mid-infrared light curve from WISE, which imaged the full sky approximately every six months in four mid-infrared bands \citep[referred to as $W1$, $W2$, $W3$, and $W4$, respectively,][]{2010AJ....140.1868W} from 2010 January to September. Since the cryogen for cooling the $W3$ and $W4$ instruments was exhausted, it was reactivated on 2013 October 3 (called NEOWISE) with the $W1$ and $W2$ bands alone \citep{2014ApJ...792...30M}. We retrieved the data of AllWISE multi-epoch photometry table and NEOWISE single-exposure (L1b) source table from the NASA/IPAC Infrared Science Archive \footnote{\url{https://irsa.ipac.caltech.edu/Missions/wise.html}}. We present the light curves in the optical and mid-infrared bands in \autoref{pic:lc}. To investigate the variability in the magnitude and color in the optical and mid-infrared, we calculated the variability in the magnitude at the ZTF zg band ($\Delta g$) and the WISE W1 band ($\Delta W1$) and the variability in the color ($\Delta (g-r)$ and $\Delta (W1-W2)$) compared to the first-epoch data.  The color variability versus magnitude variability correlation is presented in \autoref{pic:var}.

\subsection{Spectroscopy} \label{sec:spec}
We observed J1011+5442 with the Xinglong 2.16 m telescope \citep[XLT, hereafter;][]{2016PASP..128k5005F,2022ApJS..261...32F} on 2024 January 11 with a total exposure time of 4700s, the Lijiang 2.4 m telescope \citep[LJT, hereafter;][]{2019RAA....19..149W} on 2024 February 4 and 7 with the same exposure time of 3200s, and the MMT Observatory 6.5 meter telescope (MMT, hereafter) on 2024 April 5 with a total exposure time of 3000s. XLT is an equatorial mount reflector telescope at Xinglong Observatory and part of the National Astronomical Observatories of the Chinese Academy of Sciences (NAOC). The G4 grism and the slit with a width of $2\arcsec.3$ were used, so that the wavelength coverage was 4000-8800$\AA$ and the spectral resolution of $R=\lambda/ \Delta \lambda  $ was 265 at 5007 $\AA$ \citep[][]{2016PASP..128k5005F}. LJT is an altitude-azimuth mount reflector telescope at Lijiang Observatory and part of the Yunnan Observatories
(YNAO) of the Chinese Academy of Sciences (CAS). The G3 grism and the slit with a width of $2\arcsec.5$ were used, so that the wavelength coverage was 3400-9100$\AA$ and the spectral resolution of $R=\lambda/ \Delta \lambda $ was 250 at 6030 $\AA$. Binospec is a multislit imaging spectrograph that was commissioned at the MMT in 2017. The instrument we used is \textit{BINOSPEC\_270} and has a wavelength coverage of 3900-9240$\AA$ and a spectral resolution of 1340 at 6500 $\AA$ \footnote{\url{http://mingus.mmto.arizona.edu/~bjw/mmt/binospec_info.html}}. 
 J1011+5442 was observed under mostly good sky conditions by XLT, LJT, and MMT. The sky condition was slightly worse on February 4 than on February 7 for LJT. The observations by MMT were split into five 600-second exposures, one of which was made with a slightly worse seeing. 

The spectra from XLT and LJT were reduced with the pipeline \citep[{\sc PyFOSC}; ][]{yuming_fu_2020_3915021}, which is based on PyRAF for long-slit spectroscopy. The spectra from MMT were reduced with the standard pipeline \citep[{\sc PypeIt \footnote{\url{https://pypeit.readthedocs.io/en/release/index.html}}};  ][]{pypeit:joss_arXiv}. 

J1011+5442 shows type I AGN spectral features with significant broad $\rm H\alpha$ and $\rm H\beta$ emission lines based on the new spectroscopic observations. The dim-state spectra from SDSS could help us to model the stellar continuum by fitting an E-MILES SPS model \citep[][]{2016MNRAS.463.3409V} with {\sc ppxf} \citep{2004PASP..116..138C}. The stellar velocity dispersion we estimate is $123 \pm 5 {\rm \, km\, s^{-1}}$.  Based on the $M_{\rm BH}- \sigma_{*}$ relation from \citet{2002ApJ...574..740T}, we calculated the black hole mass as log($M_{\rm BH,\,\sigma_{*}}/ M_{\odot})= 7.28\pm 0.15$.

To increase the signal-to-noise ratio (S/N) of the spectra, we combined the three spectra from XLT and LJT with the tool {\texttt specutils} \footnote{\url{https://specutils.readthedocs.io/en/stable/manipulation.html}}. The spectral flux by LJT was scaled to that of XLT due to their close observational time interval within one month, which is much shorter than the typical changing-look timescale.  The absolute flux calibrations of the spectra were made using the nearest ZTF r-band magnitude with the package \texttt{pyphot} \footnote{\url{https://mfouesneau.github.io/pyphot/}}. $r=18.56\pm0.05$ at MJD 60322 and $r=18.45\pm0.05$ at MJD 60354 were adopted to scale the spectra of XLT/LJT  and MMT, respectively. The absolute flux-calibrated spectra from the new observations and the spectra from the Sloan Digital Sky Survey (SDSS) are shown in \autoref{pic:spec_ztf}.

\subsection{Spectral decomposition} \label{sec:decomp}
We further fit the spectra with a wrapper package ({\tt QSOFITMORE}; \citealt{2021zndo...5810042F}), which is based on the multicomponent spectral fitting code {\tt pyQSOFit} \citep{2018ascl.soft09008G}. The spectra were corrected for Galactic extinction
with the dust map of \cite{1998ApJ...500..525S} and the extinction law from \cite{2019ApJ...877..116W}. The dereddened spectra were then shifted to the rest frame with a redshift of $z=0.246$.

First, we decomposed the host galaxy component from the dim-state spectrum from the SDSS over the wavelength range of 3800-7000 \AA \,. We fit the spectrum with a host component, a power-law continuum, the optical \citep{1992ApJS...80..109B} and UV \citep{2001ApJS..134....1V} \feii\, template, and line components. The broad lines and Fe\,{\sc ii} emissions are weak in this spectrum \citep[see also][]{2016MNRAS.455.1691R}. The broad component of H$\alpha$ was modeled by one Gaussian profile. The stellar lines contaminate the broad components of the H$\beta$ lines \citep[][]{2016MNRAS.455.1691R}.  

Then, we fit the bright-state spectrum from SDSS and the new spectra from XLT/LJT and MMT over the same wavelength range. The broad component of H$\alpha$ and H$\beta$ was modeled by two Gaussian profiles. The host galaxy decomposition was not applied for these spectra because the negative host galaxy flux is higher than 100 pixels. We therefore subtracted the host component estimated from the dim-state spectra. For the spectrum from MMT, we masked the data in the wavelength range of 6067-6131 \AA \, because of the poor data quality.  For all the spectra, we fit the spectra with all the components as used in the dim-state spectrum. The narrow components of H$\alpha$, [N\,{\sc ii}]$\lambda\lambda$6548,6584, [S\,{\sc ii}]$\lambda\lambda$6716,6731, and H$\beta$ were each modeled by a single Gaussian profile.  The line widths and velocity offsets of the narrow lines were tied to each other. The relative flux ratio of the [N\,{\sc ii}]$\lambda\lambda$6548,6584 doublet was fixed to 2.96:1. The relative flux ratio of the [O\,{\sc iii}]$\lambda\lambda$,4959,5007 double lines was fixed to 1:3. The uncertainties were determined via a Monte Carlo simulation of the fitting procedure. We present the fitting result of the new spectra from XLT/LJT and MMT in \autoref{pic:XL} and \autoref{pic:MMT}. The properties of J1011+5442 from emission-line measurements are listed in \autoref{tab:results}. 

\begin{table*}[]
\centering
\caption{Properties of J1011+5442 estimated from different spectra.  \label{tab:results}}
\begin{tabular}{lll ccccl} \hline \hline
Spectra & $\lambda L_{\lambda}({\rm 5100 \AA})$  & FWHM of \ha\,   & FWHM of \hb\,   & $M_{{\rm BH, H}\alpha}$    & $M_{{\rm BH, H}\beta}$    & Flux of \oiii\, & reference   \\ 
        &  ( $\rm erg\,s^{-1}$)   & ( $\rm km \, s^{-1}$) & ( $\rm km \, s^{-1}$) &  ($M_{\odot}$) &  ($M_{\odot}$) &  ($\rm 10^{-17} erg \, s^{-1} cm^{-2}$) & \\ \hline
Bright state                & 6.3e43 & 2420$\pm$30   & 2350$\pm$50  & 3.6e7  & 4.7e7  & \multirow{2}*{82$\pm$3}        &  \citet{2016MNRAS.455.1691R}   \\
Dim state                   & 6.3e42 & 2720$\pm$70   & -          & 1.8e7 & -         &     &  \citet{2016MNRAS.455.1691R}    \\ \hline

Bright state  & 8.7e43 & 1621$\pm$15 & 2585$\pm$185 & 2.5e7 &  5.1e7  & 94$\pm$26   &    This work \\
Dim state          & 4.2e42 & 2824$\pm$ 0.9  & -          & 1.6e7  & -         & 91$\pm$3    & This work   \\ 
XLT/LJT           &  4.0e43 & 3248$\pm$1564   & 3036$\pm$1989 & 6.9e7  & 4.7e7 & -  &  This work   \\
MMT                    &  4.6e43 & 2409$\pm$158 & 5489$\pm$338 & 4.0e7 & 1.7e8 & 84$\pm$4  &  This work \\ 
\hline
\end{tabular}
Note: The larger error bar for the XLT/LJT spectrum is due to its poorer data quality. The BH mass is estimated based on the scaling relations for the \ha\, and  \hb\, lines from  \cite{2010ApJ...709..937G} and \cite{2006ApJ...641..689V}, respectively.
\end{table*}

\subsection{Chandra observation} \label{sec:chandra}
There are two archival \chandra\, observations (ObsID 19518 and 22525 with exposure times of 11 ks and 30 ks, respectively) of J1011+5442. Since the X-ray intensity is still weak, we only extracted the ACIS-S spectrum of ObsID 22525 (observed on 2020 September 18) with the latest {\sc ciao }(v4.16) and {\sc caldb} (v4.11.0). The source and background spectra were extracted from a circle with a radius of 3$\arcsec$ and an annulus with an inner radius of 5 $\arcsec$  and an outer radius of 10 $\arcsec$, respectively. The net photon count rate of the spectrum was only $2.6\times 10^{-3} \pm 3.3\times 10^{-4} $ photons per second. We therefore estimated the X-ray flux with the command {\it modelflux} by assuming an X-ray photon spectral index of 2 and taking the Galactic column density of hydrogen \footnote{\url{https://www.swift.ac.uk/analysis/nhtot/index.php}} to be $8.27 \times 10^{19}\,{\rm cm^{-2}}$. The lower limit of the 0.5-8 keV unabsorbed X-ray flux is $3.71\times 10^{-14} \, {\rm erg\, cm^{-2}\, s^{-1} } $, which corresponds to $L_{\rm{0.5-8 keV}} = 6.87 \times 10^{42}\, {\rm erg\, s^{-1} } $.

\section{Discussion} \label{sec:dis}
Active galactic nuclei that show repeating changing look are still rare (see also \autoref{tab:RCLAGM_lit}). We report that J1011+5442 is another repeating CLAGN after around 5 years (between 2014 and 2019) in the dim state. It returned to the type I state in 2024. The multi-epoch spectroscopic observations show significant broad \hb\, lines in 2003 and 2024 and undetectable broad \hb\, lines in 2015, while the broad \ha\, lines are always apparent. J1011+5442 is in an intermediate state between the dim state in 2015 and the bright state in 2003 based on the absolute flux-calibrated spectra (see \autoref{pic:spec_ztf}). The disappearance and reappearance of the broad \hb\, lines is consistent with the dimming and rebrightening of the optical continuum and mid-infrared emission.  The variability in color versus magnitude shows the bluer-when-brighter and redder-when-bright features in the optical and mid-infrared band, which is consistent with other CLAGNs and is not caused by the variable obscuration \citep[e.g.,][]{2018ApJ...862..109Y}. The mid-infrared (MIR) color ($W1$-$W2$) changes from 1.2 in the bright state to $0.5-0.75$ in the dim state and gradually increases as it turns on. The opposite color-change trend in the MIR band is mainly caused by the activity from the central engine of AGN (e.g., accretion rate change). Mid-IR emission is thought to mainly come from the hot dust heated by the central engine of AGNs, which lags the optical emission from the disk in the dust reverberation mapping scenario \citep{2018ARA&A..56..625H}. We estimated the time lag of the mid-infrared variability to the optical band using the interpolation cross-correlation function \citep[ICCF;][]{1998PASP..110..660P,2018ascl.soft05032S}. An interpolation time step of one day was applied to the optical and mid-infrared light curves, which were rebinned into 10 days and 180 days, respectively. The flux randomization and random subset selection methods were employed with 40,000 realizations in the Monte Carlo simulation to estimate the centroid time lag and the uncertainties \footnote{The code \texttt{pyCCF} is available in \url{http://ascl.net/code/v/1868}}. The centroid time lags for the WISE $W1$ band variability to the ZTF g and r bands are $104.7^{+18.5}_{-41.1}$ and $95.3^{+19.0}_{-29.6}$ days. According to the time lag and bolometric luminosity ($\tau$-$L$) correlation of the dust reverberation mapping from the luminous QSOs \citep{2019ApJ...886...33L}, the predicted time lags are around 165 and 62 days based on the bright- and dim-state bolometric luminosity \citep{2016MNRAS.455.1691R}. The spectroscopic observations and the optical and mid-infrared photometric light curves further support the hypothesis that the repeating changing-look phenomena are driven by the abrupt change in the intrinsic accretion rate and not by the TDE or obscuration effects \citep[e.g.,][]{2016MNRAS.455.1691R,2017ApJ...846L...7S,2018MNRAS.480.3898N}.

There are extreme variations in the broad emission lines (especially for the \hb\, lines) from the broad line region (BLR) for CLAGNs. The blue part of the spectrum from MMT is quite similar to that in the dim state from the SDSS, except for the very broad \hb\, emission lines in the MMT spectrum. The BLR might not be in equilibrium or the virial factor could change \citep[e.g.,][]{2016A&A...593L...8M,2017A&A...607L...9K}. The black hole mass (log$M_{\rm BH,\,\sigma_{*}}= 7.28\pm 0.15$) estimated based on the $M_{\rm BH}- \sigma_{*}$ relation might be a better choice \citep[e.g.,][]{2022ApJ...926..184J}, which is an intermediate value of the virial black hole mass estimated for J1011+5442 \citep[i.e., $M_{{\rm BH, bright, H}\beta}=(3.6 \pm 0.2) \times 10^7 M_{\odot}$, $M_{{\rm BH,bright, H}\alpha}=(4.7 \pm 0.1) \times 10^7 M_{\odot}$, and $M_{{\rm BH, dim, H} \alpha} =(1.8 \pm 0.3) \times 10^7 M_{\odot}$ based on the broad \hb\, and \ha\, emission lines in the bright state and \ha\, in the dim state, respectively; see][]{2016MNRAS.455.1691R}. The $M_{\rm BH,\sigma_{*}}$ is closer to the $M_{{\rm BH, dim, H} \alpha}$ in the dim state, but also consistent with the value in the bright state within the error bars. We adopted the $M_{\rm BH,\sigma_{*}}$ estimated above to estimate the Eddington ratio in different states. The Eddington ratio $L_{\rm bol}/L_{\rm EDD}=  0.28 $ was estimated for bright state with $L_{\rm bol}= 6.8 \times 10^{44}\, {\rm erg\, s^{-1} }$ \citep[][]{2016MNRAS.455.1691R}. When we assume a bolometric correction factor of 10 \citep[e.g.,][]{2009MNRAS.399.1553V}, the estimated Eddington ratio from the X-ray observation for J1011+5442 is $L_{\rm bol}/L_{\rm EDD} \gtrsim 2.8 \% $. This is similar to but slightly lower than that ($L_{\rm bol}/L_{\rm EDD}=  3.5\% $) of the dim state with $L_{\rm bol}= 8.4 \times 10^{43}\, {\rm erg\, s^{-1} }$ \citep[][]{2016MNRAS.455.1691R}, which suggests that J1011+5442 is still in the dim state at the epoch (2020) of the X-ray observation. The host-subtracted bolometric luminosity estimated from the MMT spectrum (2024) is $L_{\rm bol}=5.2 \times 10^{44}\, {\rm erg\, s^{-1} }$. The corresponding Eddington ratio of $L_{\rm bol}/L_{\rm EDD}$ is 0.21, which indicates that J1011+5442 is in the intermediate state between the dim (2015-2020) and bright state (2003-2010). The Eddington ratio (2.8 \% - 28\%) is close to the critical value at which the state transition between ADAF and SSD would occur \citep[e.g.,][]{2018MNRAS.480.3898N,2021MNRAS.508..144G}. The accretion mode transition could explain the significant change in the broad \hb\, lines.

The changing-look timescale for the disappearance and reappearance of broad \hb\, lines in J1011+5442 is around 5 years (see \autoref{pic:lc}). The well-known CLAGN Mrk 1018 also experienced a changing look multiple times \citep[e.g.,][]{2016A&A...593L...8M,2017A&A...607L...9K,2021MNRAS.506.4188L,2023A&A...677A.116B}, which is somewhat similar to the behavior of J1011+5442 (e.g., the changing-look timescale and the variability in the mid-infrared emission). The mechanism that drives the variation in the accretion rate on this short timescale with multiple changing-look events in individual sources is still debated. For a recoiling supermassive black hole, the sound crossing time ($\tau_\mathrm{s}$) of the density perturbation in the accretion disk caused by a tidal impulse is given by \citep{2008ApJ...677..884L,2018ApJ...861...51K} 
\begin{equation}
\centering
\tau_\mathrm{s} \sim 70 \, (\frac{M_\mathrm{BH}}{10^8M_{\odot}})(\frac{R_d}{1000 R_g}) (\frac{T}{10^5 K})^{-1/2} \, \mathrm{years},
\end{equation}  
where the effective temperature of the accretion disk $T$ is determined by 
\begin{equation}
\centering
T = 10^{5.56} (\frac{L_\mathrm{bol}}{L_{\rm Edd}})^{1/4}(\frac{M_{\rm BH}}{10^8M_{\odot}})^{1/4} \, \mathrm{K}
\end{equation}  
for a rotating Kerr black hole. We obtain that the $\tau_\mathrm{s}$ is around $1.0\sim 1.4$ years adopting a typical accretion disk radius ${R_d}\sim 100 R_g$ \citep[e.g.,][]{2018MNRAS.480.3898N}, log($M_{\rm BH,\,\sigma_{*}}/ M_{\odot})= 7.28$, and the bolometric Eddington ratio $0.028 \sim 0.28$. The sound crossing timescale is much shorter than the observed changing-look timescale of 5 years, which does not support a recoiling supermassive black hole in J1011+5442. In the rSMBH scenario, \citet{2018ApJ...861...51K} predicted that Mrk 1018 would return to type I in the mid-2020s. It experienced an outburst in the optical and mid-infrared bands in 2021 \citep[e.g.,][]{2023A&A...677A.116B}. However, this scenario is not supported by the new spectropolarimetry observation with low continuum polarization signatures for Mrk 1018\citep[e.g.,][]{2020A&A...644L...5H}. A low polarization degree is also reported for J1011+5442 \citep[][]{2017A&A...604L...3H,2019A&A...625A..54H}, which further rules out the rSMBH explanation for the repeating changing-look phenomena in J1011+5442 \citep[e.g.,][]{2020A&A...644L...5H}. 

In the scenario of close binaries of supermassive black holes, the tidal interaction between mini-disks and CB-SMBHs with low mass ratios and high eccentricities could lead to the turn-on and turn-off due to the accumulation/exhaustion of peeled gas after the periastron phase \citep[e.g.,][]{2020A&A...643L...9W}. The asymmetric or double-peaked profiles in the broad emission lines (see \autoref{pic:hbha}) might support this scenario. The discovery of more repeating CLAGNs can also provide us with candidate SMBHBs.


\section{Conclusion} \label{sec:con}
We reported that J1011+5442 is another repeating CLAGN based on new spectroscopic observations by XLT, LJT, and MMT in 2024. The broad \hb\, lines in J1011+5442 disappeared and reappeared within around 20 years. J1011+5442 gradually returns to the type I state, which is located in between the bright state in 2003 and the dim state in 2015. The rapid variations in the broad Balmer lines are accompanied by the increase in the optical continuum and in the mid-infrared emission since 2019. The rebrightening of its optical and mid-infrared emission supports the hypothesis that the changing-look phenomenon is driven by the abrupt change in the accretion rate because the time lag of the mid-infrared emission variability relative to the optical band is consistent with the prediction of dust reverberation mapping. Multiple times of changing-look events within a short timescale are still rare and confusing. We encourage more optical spectroscopic and X-ray observations to investigate the repeating changing-look behavior for J1011+5442 and to determine whether the broad \hb\, line again disappears in the next several years.

\begin{acknowledgements}
We thank the anonymous referee for useful comments that effectively improved the manuscript. We are thankful for the support of the National Science Foundation of China (11721303, 11927804, and 12133001) and the National Key R \& D Program of China (2022YFF0503401). We acknowledge the science research grant from the China Manned Space Project with No. CMS-CSST-2021-A06. We acknowledge the support of the staff of the Xinglong 2.16m telescope. This work was partially supported by the Open Project Program of the Key Laboratory of Optical Astronomy, National Astronomical Observatories, Chinese Academy of Sciences.

Guoshoujing Telescope (the Large Sky Area Multi-Object Fiber Spectroscopic Telescope LAMOST) is a National Major Scientific Project built by the Chinese Academy of Sciences. Funding for the project has been provided by the National Development and Reform Commission. LAMOST is operated and managed by the National Astronomical Observatories, Chinese Academy of Sciences. We acknowledge the support of the Cultivation Project for LAMOST Scientific Payoff and Research Achievement. 

We acknowledge the support of the staff of the Xinglong 2.16m telescope and the Lijiang 2.4m telescope, which was partially supported by the Open Project Program of the Key Laboratory of Optical Astronomy, National Astronomical Observatories, Chinese Academy of Sciences and funded by the Chinese Academy of Sciences and the People’s Government of Yunnan Province, respectively. Observations reported here were obtained at the MMT Observatory, a joint facility of the University of Arizona and the Smithsonian Institution.

Based on observations collected at the Samuel Oschin Telescope 48-inch and the 60-inch Telescope at the Palomar Observatory as part of the \textit{Zwicky} Transient Facility project. The ZTF forced-photometry service was funded under the Heising-Simons Foundation grant 12540303 (PI: Graham). This publication makes use of data products from the Wide-field Infrared Survey Explorer and NEOWISE, which is funded by the National Aeronautics and Space Administration. This research has made use of data products from the Sloan Digital Sky Survey (SDSS, \url{www.sdss.org}), and NASA/IPAC  EXTRAGALACTIC  DATABASE (NED, \url{https://ned.ipac.caltech.edu/}). 
\end{acknowledgements}

%
%
\bibliographystyle{aa}
\bibliography{ref}

\begin{appendix}
\onecolumn
\section{Fitting}
\begin{figure*}
\centering
\includegraphics[width=0.6\textwidth]{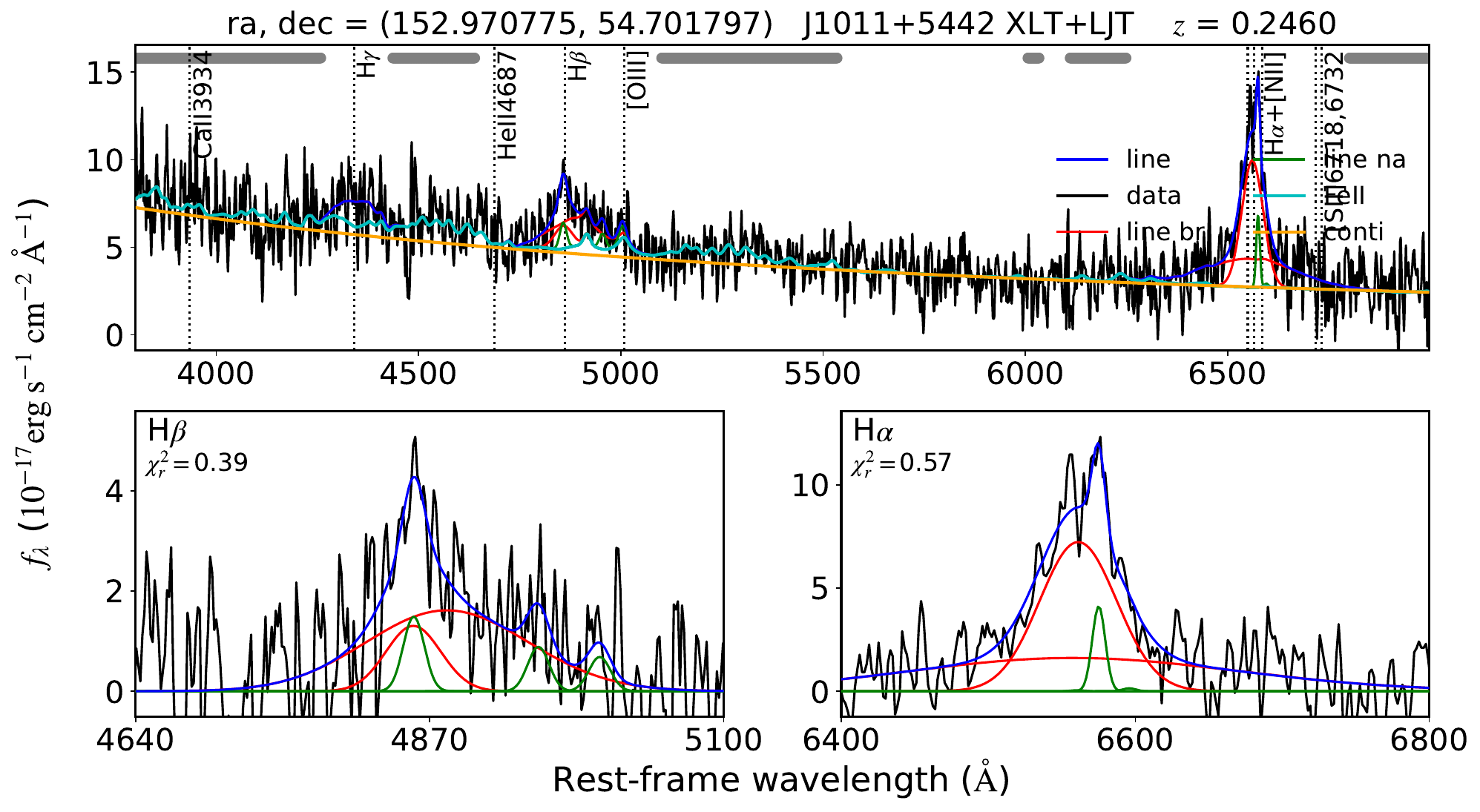}
\caption{Spectral fitting result for the combined spectrum of J1011+5442 from XLT and LJT. }
\label{pic:XL}
\end{figure*}

\begin{figure*}
\centering
\includegraphics[width=0.6\textwidth]{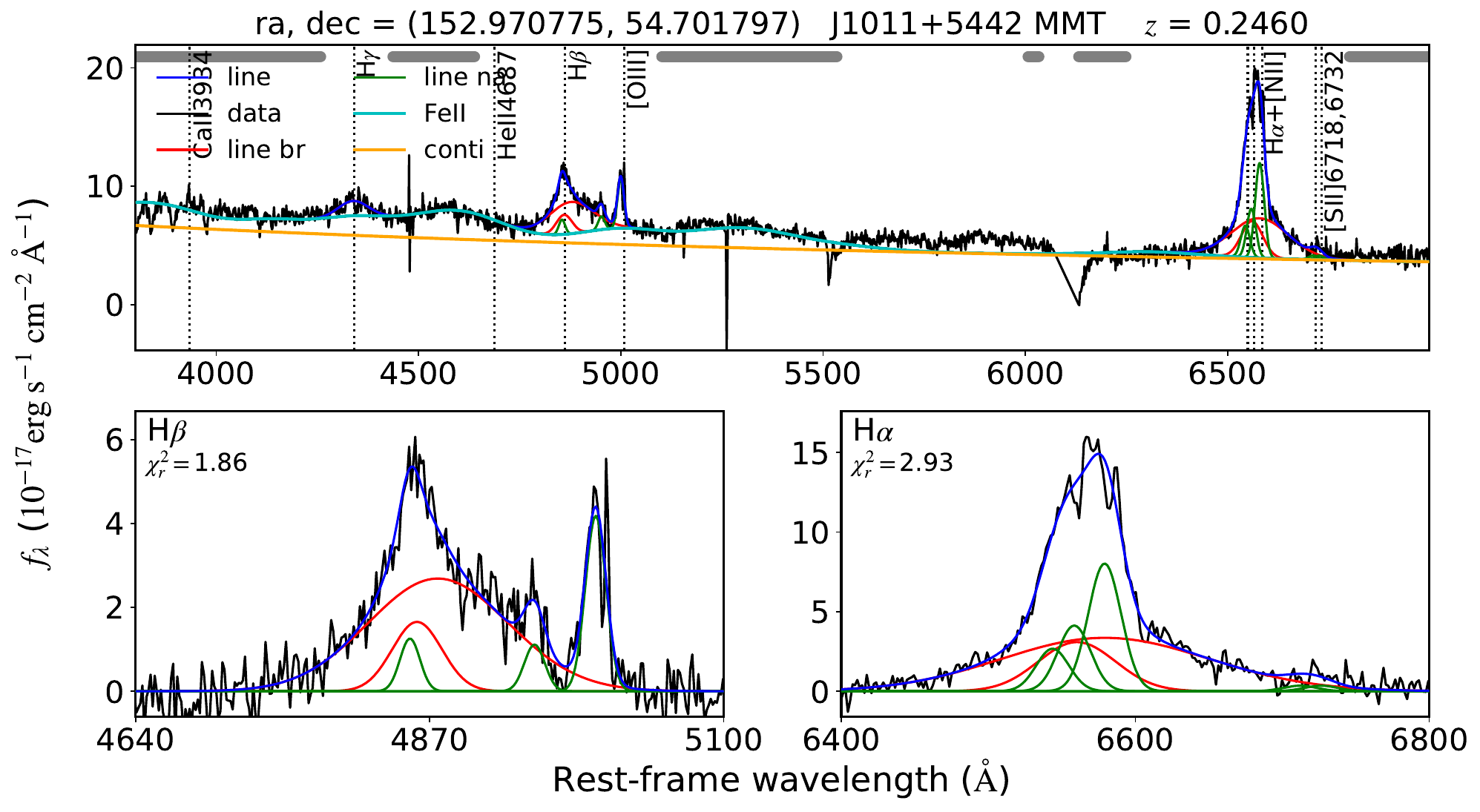}
\caption{Spectral fitting result for the spectrum of J1011+5442 from MMT. }
\label{pic:MMT}
\end{figure*}

\section{RCLAGNs}
\begin{table*}
\scriptsize
\caption{Repeating CLAGNs collected from the literature. The table lists the source name, redshift, BH mass, types, and references. \\  \label{tab:RCLAGM_lit}}
\centering
 \begin{tabular}{lc cc cc} \hline\hline
{Name} &   {Redshift}&  {log($M_\mathrm{BH}/M_{\odot}$)}  & {Type} & {Ref} & \\  \hline
3C 390.3 &  0.05613 &  9.3 &  ﻿1 (1975) -> 1.9 (1980-1984) -> 1 (1985-1988) -> 1 (2005-2014)  &  (1) &  \\
B3 0749+460A &  0.05174 &  8.11 &  1.9(2004)->1.8(2013)->1.9(2016-2021) &  (2) &  \\
Fairall 9 &  0.04614 &  8.41 &  ﻿1 (1977-1981) -> 1.8 (1984) -> 1 (1987) &  (1) &  \\
Mrk 1018 &  0.04296 &  7.84 &  ﻿1.9 (1979) -> 1 (1984-2009) -> 1.9 (2015) &  (3) &  \\
Mrk 590 &  0.02638 &  7.5 &  ﻿1.5 (1973) -> 1 (1989-1996) -> 1.9 (2006-2014) ->1 (2017)  &  (4) &  \\
NGC 1365 &  0.00546 &  6.65 &  1(1993)->1.9(2009-2010)->1(2013-2014-2017)-2(2021) &  (5,6) &  \\
NGC 2992 &  0.00771 &  7.72 &  1(1977-1979)->2(1985-1994)->1(1998-2001)->2(2006-2007)->(2014-2021) &  (7) &  \\
NGC 1566 &  0.00502 &  6.92 &  2(1970)->1(1985)->1.8(2013)->1.2(2018)->1.8(2019) &  (8,9) &  \\
﻿NGC 3516 &  0.00884 &  7.49 &  ﻿1 (1996-1998) -> 1 (2007) -> 2 (2014-2017) &  (1) &  \\
NGC 4151 &  0.00333 &  7.56 &  ﻿1 (1974) -> 1.9 (1984-1989) -> 1.5 (1990-1998) -> 1.8 (2001) &  (10,1) &  \\
NGC 5548 &  0.01717 &  7.51 &  1(1998)->1.8(2007)->1(2009) &  (11,12) &  \\
NGC 7582 &  0.00525 &  7.74 &  ﻿2 (1980-1998) -> 1 (1998)->1.8(2004)  &  (13,1) &  \\
NGC 7603 &  0.02876 &  8.11 &  ﻿1 (1974) -> 1.8 (1975) -> 1 (1976-1998)  &  (1) &  \\
﻿SDSS J0917272-645628 &  0.08600 &  $-$ &  1.9(2005)->1.5(2017)->1.8(2019)->1.9(2020) &  (14) &  \\
SDSS J132457.29+480241.3 &  0.27156 &  8.4 &  1(2003)->1.9(2014)->1.8(2021)->1(2022) &  (15,16) &  \\
﻿SDSS J1340153-045332 &  0.08654 &  $-$ &  1.5(2005)->1.9(2020.02)->1.5(2020.06) &  (14) &  \\
SDSS J151652.48+395413.4 &  0.06323 &  7.57 &  1(2003)->2(2021)->1.9(2023) &  (17) &  \\
SDSS J162829.17+432948.5 &  0.25994 &  8.2 &   1(2001.05)->1.5(2016.02)->1.9( 2021.04-2021.06)->1.5(2021.08-2022.05) &  (18,15) &  \\
UGC 3223 &  0.01562 &  7.99 &  1.5(﻿1987) -> 2(2013-2014) -> 1.8(2020)  &  (19) &  \\ \hline \\
\end{tabular}\\
Notes: (1)~\citet{2019sf2a.conf..509M} (2)~\citet{2022RAA....22a5011W} (3)~\citet{2016A&A...593L...8M} (4)~\citet{2014ApJ...796..134D} (5)~\citet{2014ApJ...795...87B} (6)~\citet{2023MNRAS.518.2938T} (7)~\citet{2021MNRAS.508..144G} (8)~\citet{2019MNRAS.483L..88P} (9)~\citet{2020MNRAS.498..718O} (10)~\citet{2007MNRAS.377..607P} (11)~\citet{2016ApJS..225...29B} (12)~\citet{2020A&A...641A.167S} (13)~\citet{2018MNRAS.473.5334R} (14)~\citet{2022MNRAS.511...54H} (15)~\citet{2024ApJ...970...85W} (16)~\citet{2024ApJ...966...85Z} (17)~\citet{2023ApJ...956..137W} (18)~\citet{2022ApJ...939L..16Z} (19)~\citet{2020ApJ...901....1W} 
\end{table*}

\clearpage
\end{appendix}





\end{document}